\documentclass[pdflatex,sn-nature]{sn-jnl}

\makeatletter
\providecommand{\ref@jnl}[1]{#1} 
\makeatother

\usepackage[numbers]{natbib}  

\usepackage{graphicx}%
\usepackage{multirow}%
\usepackage{amsmath,amssymb,amsfonts}%
\usepackage{amsthm}%
\usepackage{mathrsfs}%
\usepackage[title]{appendix}%
\usepackage{xcolor}%
\usepackage{textcomp}%
\usepackage{manyfoot}%
\usepackage{booktabs}%
\usepackage{algorithm}%
\usepackage{algorithmicx}%
\usepackage{algpseudocode}%
\usepackage{listings}%

\theoremstyle{thmstyleone}%
\theoremstyle{thmstyletwo}%

\theoremstyle{thmstylethree}%

\begin{document}


\title[Article Title]{The Search for Intermediate-Mass Black Holes in the Time Domain Era}

\title[Article Title]{Variability as a new discovery channel for Intermediate-Mass Black Holes in the Time Domain Era}



\author*[1]{\fnm{Colin J.}\sur{Burke}}\email{colin.j.burke@yale.edu}

\author[1,2,3]{\fnm{Priyamvada} \sur{Natarajan}}\email{priyamvada.natarajan@yale.edu}

\affil*[1]{\orgdiv{Department of Astronomy}, \orgname{Yale University}, \orgaddress{\street{219 Prospect Street}, \city{New Haven}, \state{CT}, \postcode{06511}, \country{USA}}}

\affil[2]{\orgdiv{Department of Physics}, \orgname{Yale University}, \orgaddress{\street{217 Prospect Street}, \city{New Haven},  \state{CT}, \postcode{06511}, \country{USA}}}

\affil[3]{\orgdiv{Black Hole Initiative}, \orgname{Harvard University}, \orgaddress{\street{20 Garden Street}, \city{Cambridge},  \state{MA}, \postcode{02138}, \country{USA}}}


\abstract{Between the groundbreaking detections of stellar-mass black holes by LIGO/Virgo/KAGRA and JWST’s revelation of a surprisingly abundant population of supermassive black holes, one crucial missing link remains: the elusive intermediate-mass black holes (IMBHs). IMBHs represent a key phase in the hierarchical growth of black holes, yet they have persistently evaded detection. Traditional methods, effective for both actively accreting and quiescent black holes, have largely failed to uncover this hidden population. Here, we argue that novel observational strategies—particularly time-domain variability studies of active galactic nuclei (AGN) and tidal disruption events—provide a promising path forward. Finding IMBHs will resolve critical gaps in our understanding of black hole formation and the various mechanisms driving their subsequent growth. The upcoming Vera C. Rubin Observatory, with its unprecedented capacity to monitor the dynamic sky, stands to revolutionize our ability to detect these long-sought IMBHs, shedding new light on the assembly history of black holes across cosmic time.}

\maketitle

\section{Introduction}\label{sec:intro}

Evidence supporting the existence of supermassive black holes (SMBHs) with masses $M_{\rm{BH}} \gtrsim 10^6 M_{\odot}$ harboured in the centers of nearly every nearby galaxy has been accumulating for over thirty years \citep{Magorrian1998,Kormendy1995}. However, the population of intermediate-mass black holes (IMBHs) with masses $M_{\rm{BH}} \sim 100 - 100,000 M_{\odot}$ remains elusive despite being an inevitable pre-cursor phase in the growth history of SMBHs \citep{Greene2020}. To explain the ubiquity of SMBHs discovered at high redshifts of $z\sim9{-}11$ when the universe was only several hundred Myrs old and the properties of accreting sources like ULAS J1342+0928 \citep{Banados2018}; UHZ1 \citep{Natarajan2024}; GHZ9 \citep{Kovacs2024}, GNz11 \citep{Maiolino2024}, it appears that these SMBHs, in place so early in the Universe, likely grew from ``heavy'' IMBH seeds of $M_{\rm{BH}} \sim 10^{3-5} M_{\odot}$, or from rapid super-Eddington accretion onto ``light'' IMBH seeds of $M_{\rm{BH}} \sim 10^{1-2} M_{\odot}$, or from a combination thereof \citep{LodatoPN2006,LodatoPN2007,  Volonteri2008,Bellovary2011,Greene2012,Alexander2014,Natarajan2014,Inayoshi2020}.

Since black holes themselves emit no light, detection methods to date rely on either their gravitational effects or electromagnetic radiation emitted by the hot material in their surrounding accretion disks. The former approach has enabled studies from stellar or gas dynamical modeling when the black hole sphere of influence can be spatially resolved (e.g., \citep{Gerssen2002,Gebhardt2005,Noyola2010,Lutzgendorf2013,Lutzgendorf2015,denBrok2015,Baumgardt2017,Kiziltan2017,Nguyen2019}), and from signatures from hyper-velocity stars (e.g., \cite{Edelmann2005,Haberle2024,Han2025}). The latter approach has enabled the detection of SMBHs in the centers of galaxies that manifest as active galactic nuclei (AGN). The sphere of influence of a typical IMBH is incredibly compact and ranges from a fraction of a parsec to a few parsecs making it observationally unresolvable even with VLBI at LIGO source distances. Meanwhile, IMBHs have tidal radii larger than their gravitational radii, making them more suitable for triggering detectable tidal disruption events (TDEs) from stars that stray close. Because the masses and stellar velocity dispersions of galaxies correlate with the mass of their central black holes in the local Universe (e.g., \citep{Merritt2001,Tremaine2002,Kormendy2013,Reines2013}), low-mass systems---dwarf galaxies with stellar masses below $10^{10}\,M_\odot$---have become prime targets in the search for IMBHs. A well-known local example, NGC 4395, has been studied for more than 20 years and remains one of the lowest-mass SMBHs known in a dwarf galaxy, with a black hole mass of $\sim 1\times10^4 - 2\times10^5 M_{\odot}$  \citep{Filippenko2003}. On the other end of the mass spectrum, the groundbreaking detections of gravitational waves from coalescing stellar-mass black holes by LIGO/Virgo/KAGRA have revealed sources like GW190521, with a remnant mass of $142^{+28}_{-16} M_{\odot}$ \citep{Abbott2020} and GW231123 in the recently completed O4 Run with a final remnant mass in excess of $200 M_{\odot}$  \citep{LIGO+2025}. 

The Laser Interferometer Space Antenna (LISA) is destined to revolutionize our understanding of the IMBH population to $z\sim 20$ and the merger-driven growth channel of SMBHs \citep{Colpi2017}. However, dynamics will complicate the interpretation of LISA events and the subsequent inference of seed formation mechanisms \cite{Ricarte2018}. Depending on seed formation pathways, IMBHs might exist in a variety of cosmic environments: in the centers of dwarf galaxies; in dense star clusters; as wanderers in galactic halos; or as gravitationally bound IMBH-SMBH pairs known as intermediate mass ratio inspirals. Electromagnetic observations of the accretion properties and local environments of IMBHs will be essential to gather a more complete picture of black hole mass assembly over cosmic time. A multi-messenger approach is needed to close the current gaps in our understanding of the formation, fueling, and feedback from these black hole populations.

In this \emph{perspective} article, we review some recent advances in identifying IMBHs and make the case for leveraging time-domain accretion signatures for further discoveries. Specifically, we advocate for wide-field searches for AGN variability and flares from tidal disruption events as extremely promising avenues for identifying IMBHs in their diverse environments. In \S\ref{sec:snapshot}, we briefly review the non-time-varying (``snapshot'') methods for detecting accreting IMBHs, such as broad emission line detections in large spectroscopic surveys and detections from deep X-ray and photometric surveys. For a more comprehensive review of these ``snapshot" methods, see \citep{Greene2020,Mezcua2017} and the \emph{perspective} piece by \cite{Reines2013}. In \S\ref{sec:var}, we discuss recent exciting progress in identifying IMBHs using time-domain accretion signatures, namely AGN variability and TDE flares. In \S\ref{sec:prosp}, we present new methods for inferring the properties of IMBHs using variability and how to leverage it to forecast detection strategies for the population. There are however, challenges with these techniques as well and they are discussed in \S\ref{sec:challenges}. We conclude in \S\ref{sec:concl}, and advocate for the upcoming Rubin Legacy Survey of Space and Time (LSST) as the discovery machine for IMBHs via TDE flares and dwarf AGN variability. 

\begin{figure*}[ht]
\centering
\includegraphics[width=0.95\textwidth]{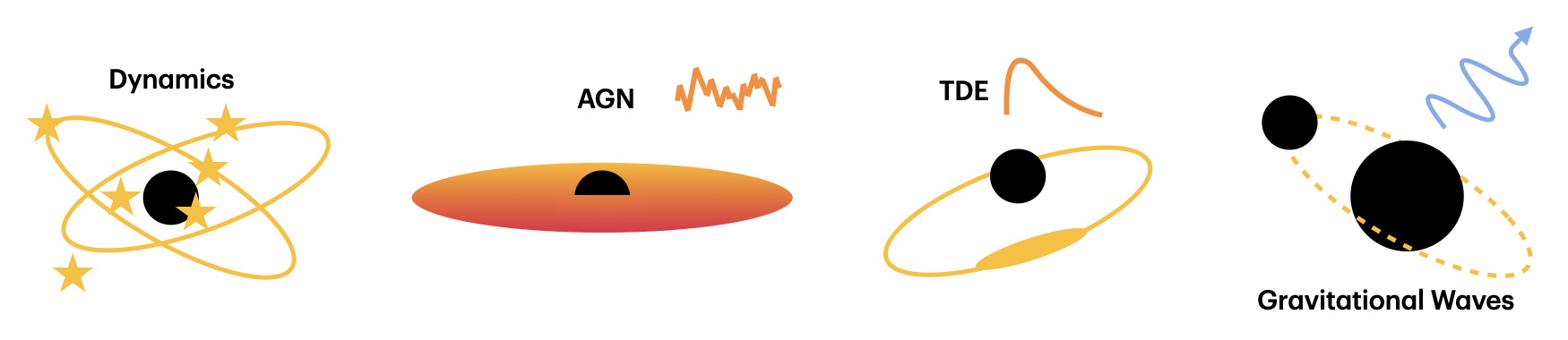}
\caption{\textbf{Schematic showing the primary methods for identifying IMBHs.} Left to right: Dynamical signatures, AGN accretion, TDE flares, and gravitational waves.}\label{fig:cartoon}
\end{figure*}

\section{``Snapshot'' methods for detecting IMBHs}\label{sec:snapshot}

\begin{figure*}[ht]
\centering
\includegraphics[width=0.95\textwidth]{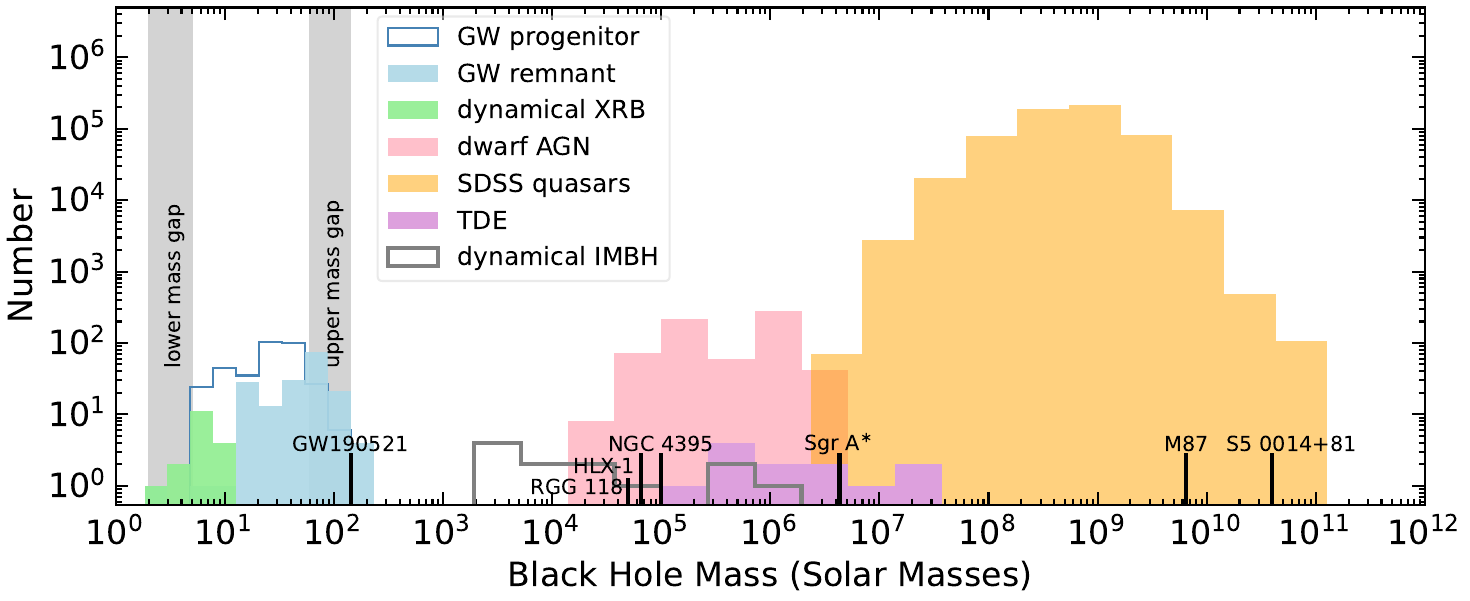}
\caption{\textbf{The observed black hole mass spectrum and the IMBH ``mass desert''.} Included in the figure are stellar-mass black holes with dynamical mass measurements in X-ray binary (XRB) systems \citep{Corral-Santana2016}; confidently-detected gravitational wave (GW) events \citep{TheLIGOScientificCollaboration2025}; dynamical IMBH detections \citep{Gerssen2002,Gebhardt2005,Noyola2010,Lutzgendorf2013,Lutzgendorf2015,denBrok2015,Baumgardt2017,Kiziltan2017,Nguyen2019,Haberle2024}; supermassive black holes from SDSS quasars with single-epoch black hole mass estimates \citep{Wu2022}; broad-line dwarf AGNs with single-epoch black hole mass estimates \citep{Greene2007,Dong2012,Reines2013,Liu2018,Chilingarian2018}; and tidal disruption events (TDEs) with masses estimated from the stellar velocity dispersion of their host galaxy bulges \citep{Wevers2017}. The lower mass gap refers to the scarcity of black holes below $\sim 5 M_\odot$ and above the maximum possible mass of a neutron star (e.g., \citep{Kreidberg2012}). The upper mass gap results from predictions from stellar evolution that pair instability supernova leave no remnants above $\sim 140 M_\odot$ (e.g., \citep{Woosley2017}). We mark some well known black holes: GW190521 \citep{Abbott2020}; RGG 118 \citep{Baldassare2015}; HLX-1 \citep{Farrell2009}; NGC 4395 \citep{Filippenko2003}; Sgr A$^{\ast}$ \citep{EventHorizonTelescopeCollaboration2022}; M87 \citep{EventHorizonTelescopeCollaboration2019}; and the luminous blazar S5 0014+81 \citep{Kuhr1983,Ghisellini2009}.\label{fig:mass}}
\end{figure*}

In this section, we briefly review some traditional methods that have been deployed for detecting accreting IMBHs excluding gravitational waves and kinematic signatures (see Figure~\ref{fig:cartoon}). We show an illustrative histogram of black hole mass measurements collated from the literature in Figure~\ref{fig:mass}, ranging from stellar mass black holes in X-ray binary systems \citep{Corral-Santana2016} to SMBHs in Sloan Digital Sky Survey (SDSS) quasars \citep{Wu2022} to make the point that there is a lack of observed black hole mass measurements in the IMBH ``mass desert.''

IMBHs are currently probed primarily through dynamical and accretion-based methods. Stellar and gas kinematics in globular clusters and nuclear star clusters can reveal dark central masses, while ultra- and hyperluminous X-ray sources, low-luminosity AGN in dwarf galaxies, and compact radio cores interpreted via the “fundamental plane” of black-hole activity highlight actively accreting IMBH candidates. Planned future facilities expected to come online in the coming decades are expected sharpen these measurements. For instance, Thirty–meter class telescopes with adaptive optics will enable routine dynamical measurements of IMBH spheres of influence in nearby systems, and next-generation X-ray and radio surveys (e.g., ngVLA/SKA-class instruments) will extend sensitivity to faint, low-Eddington accretion sources. 

\subsection{Spectroscopic Identification}

Motivated by individual examples of broad-line dwarf AGNs such as NGC 4395 and Pox 52 \citep{Barth2004}, \cite{Greene2007} performed a systematic search for AGN spectroscopic signatures of AGNs with broad-line black hole masses $< 2{\times}10^{6} M_{\odot}$ in the SDSS. The primary spectroscopic AGN signatures are broad emission lines; narrow line ratio tests; and coronal lines. The sample of spectroscopically-selected dwarf AGNs has since expanded to a few thousand \citep{Reines2013,Chilingarian2018,Mezcua2024,Pucha2025}. However, we caution that authors have used different upper thresholds for the galaxy stellar mass ranging from $M_{\star} \sim 3\times 10^9 M_{\odot}$ to $2\times10^{10} M_{\odot}$ in deriving these estimates. A strength of spectroscopic selection is that the black hole mass can be directly estimated from the broad lines \citep{Shen2013}. However, broad lines can also be created by stellar sources, such as supernova remnants or stellar-driven winds, without the presence of an underlying AGN (e.g. \citep{Izotov2007}). \cite{Baldassare2016} found that 14/16 of the \cite{Reines2013} broad-line AGNs had ambiguous or transient broad emission. Besides, IMBHs with masses below $M_{\rm{BH}} \sim 10^5 M_{\odot}$ or Eddington ratios below $\sim 0.01$ might not photo-ionize the broad line region at all \citep{Chakravorty2014,Kang2025}. Additionally, it is now understood that narrow emission line ratios may not reveal the presence of AGNs in low-metallicity galaxies \citep{Groves2006,Cann2019} and may instead indicate vigorous star formation. Future prospects for AGN detection are however, extremely promising, with ongoing spectroscopic surveys such as SDSS-V, DESI, and 4MOST that will no doubt substantially expand the total number of dwarf AGNs.

\subsection{Color and SED Selection}

AGNs can be identified from their blue optical colors, indicating a power-law continuum, or from their mid-infrared properties that trace the presence of a dusty torus (e.g., \citep{Hviding2022}). However, in dwarf galaxies, these tracers are dominated by recent star formation activity and the properties of galactic dust rather than AGN activity. For example, it has been demonstrated that mid-infrared selection from the WISE survey suffers from strong contamination from low-redshift starburst galaxies (e.g., Figure~1 of \citep{Hainline2016}).

\subsection{Radio and X-ray Searches}

Radio and X-ray emission are a nearly ubiquitous feature of black hole accretion. X-ray surveys have uncovered several dwarf AGNs \citep{Birchall2020,Mezcua2018,Sacchi2024}. Unlike searches in the optical, those in the radio and X-rays are independent of dilution from host galaxy starlight or photo-ionization. However, X-ray detections can be contaminated by X-ray binary populations, ultraluminous X-ray sources, and background quasars. Furthermore, beyond the eROSITA survey, deep and homogeneous surveys are currently unavailable in X-ray wavelengths. Meanwhile, radio emission can also be produced by star formation processes \citep{Condon1992}. Radio and X-ray variability can be used to differentiate between IMBHs and star formation. The difference in their variability properties or the fundamental plane (e.g., \citep{Gultekin2019}) could also be used to distinguish between variability from an IMBH, X-ray binaries, or accreting neutron stars that are known to power at least some ultraluminous X-ray sources \citep{Lasota2011}. However, it is possible that some AGNs lack X-ray emission or present very weak X-ray emission, as expected for super-Eddington accretion (e.g., \citep{Arcodia2024}). X-ray weakness is seen in some nearby compact emission line galaxies \citep{Simmonds2016}, and is believed to be the case for the high redshift little red dots detected by the James Webb Space Telescope (JWST) \citep{Inayoshi2024,Pacucci2024,Ananna2024}.

\section{New Time-Domain Approaches}\label{sec:var}

\subsection{Dwarf AGN Variability}

Stochastic variability on all timescales is a ubiquitous feature of AGNs (e.g., \cite{Ulrich1997}). Motivated by alternate approaches to identify dwarf AGNs that were missed by previous searches, \cite{Baldassare2018} identified a sample of 135 AGN candidates at $z<0.15$ from variability using difference imaging from SDSS images of Stripe 82, of which 35 sources have $M_{\star}<10^{10} M_{\odot}$. Notably, only 86 (82) of these candidates were confirmed by narrow emission line ratio (broad-line width) tests. The sample of local dwarf AGN candidates was later expanded using the Palomar Transient Factory \citep{Baldassare2020}; the Zwicky Transient Facility \citep{Ward2022,Bernal2024}; and other surveys \citep{Martinez-Palomera2020,Wasleske2022}. \cite{Eberhard2024} identified 10 dwarf AGNs with $M_{\star}<3{\times}10^{9} M_{\odot}$ from the Very Large Array Sky Survey, 8 of which were found to be variable in the radio. The surprisingly large number of dwarf AGN candidates with stellar masses of $M_{\star} \sim 10^{8.5}-10^{10} M_{\odot}$ implies a high black hole occupation fraction in this stellar mass range and the existence of a large fraction of dwarf AGNs missed by other selection methods. Fortunately, the long-duration (months$-$decades) and stochastic nature of AGN variability is very different from typical supernova variability, allowing them to be cleanly differentiated \citep{Butler2011}. Furthermore, variability of the little red dots is an important AGN diagnostic (e.g., \citep{Tee2024,Furtak2025}). However, some variable stars (e.g., luminous blue variables, cataclysmic variables) can also closely mimic AGN variability.

Using $\sim$ weekly imaging from the Dark Energy Survey supernova program in three deep fields, \cite{Burke2022des} identified 26 candidate dwarf AGNs with $M_{\star} < 10^{9.5} M_{\odot}$ out to $z\sim1$ using high-precision light curves from deep single-epoch imaging. A population of low-mass variable AGNs was also identified in the COSMOS field from Hyper Suprime Camera supernova search program \citep{Kimura2020}. Like deep X-ray and radio searches, these higher redshift studies are complicated by uncertainties in photometric redshifts and stellar mass estimates in sources with significant emission from both star formation and AGN. If the AGN emission swamps the host galaxy component entirely, the scatter in the stellar mass estimated from spectral energy distribution modeling can be more than an order of magnitude \citep{Ciesla2015,Sanchez-Saez2019,Burke2024hsc}. Going forward, fully Bayesian approaches will be critical to mitigating the degeneracies in AGN-host decomposition \citep{Buchner2024}.

\begin{figure*}[ht]
\centering
\includegraphics[width=0.95\textwidth]{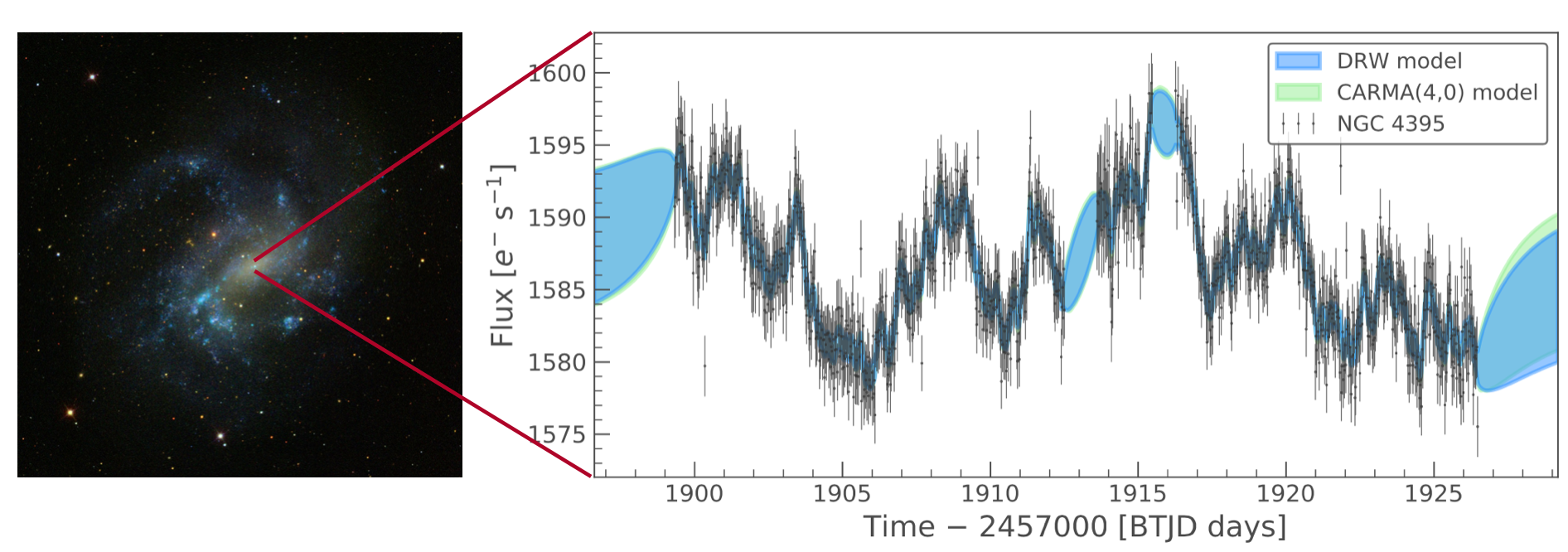}
\caption{\textbf{$\sim$ 1 month-long TESS optical light curve of the nucleus of the dwarf AGN NGC 4395.} Left: SDSS color image of NGC 4395, that contains one of the lowest mass SMBHs known. Right: The TESS light curve (black points with 1$\sigma$ uncertainties) and fitted with a damped random walk (DRW; blue) or higher-order continuous autoregressive moving average (CARMA; green) process \citep{Kelly2014}. The blue and green bands mark the 1$\sigma$ uncertainty range.} Color image credit: Sloan Digital Sky Survey.\label{fig:4395}
\end{figure*}

\subsubsection{NGC 4395 as the archetypal variable dwarf AGN}

NGC 4395 has long been known to be highly variable on time scales of hours to days in both optical and X-ray \citep{Lira1999,Iwasawa2000,Vaughan2005}. Using high-cadence imaging from the Transiting Exoplanet Survey Satellite (TESS), \cite{Burke2020TESS} extracted the optical light curve of NGC 4395, shown in Figure~\ref{fig:4395}. The variability is well described by a damped random walk process with a damping timescale of a few days \citep{Burke2020TESS}, although a steeper PSD (Power Spectral Density) than expected from a damped random walk process is observed on short timescales of less than a day \citep{Beard2025}. Despite the high cadence and photometric precision of TESS, the very poor angular resolution of nearly an arcminute results in a large fraction of host light diluting the variability signal. Therefore, AGN variability studies using TESS are mostly limited to highly variable sources such as NGC 4395 or bright blazars. However, these results imply that high-precision photometry with a typical cadence of $\sim$ hours-days can reveal the variability of accreting IMBHs.

The black hole mass estimate of NGC 4395 has an uncertainty of about an order of magnitude, ranging from $M_{\rm{BH}} \sim 1\times10^4 - 2\times10^5 M_{\odot}$ \citep{Pandey2024}. Dynamical measurements and reverberation mapping using the C~IV line fall at the upper end of this range ($M_{\rm{BH}} \sim 10^5 M_{\odot}$) \citep{Peterson2005,denBrok2015}. However, recent H$\alpha$ reverberation mapping measurements fall on the lower end of the measurement range ($M_{\rm{BH}} \sim 10^4 M_{\odot}$) \citep{Cho2021}. Systematic errors from line profile modeling, AGN continuum, host galaxy contamination, and the geometric virial factor might explain some of these differences. Its black hole mass is controversial and represents an embarrassing gap in our knowledge, specially if we wish to use NGC 4395 as a cornerstone to anchor low-mass end of the BH mass- stellar mass scaling relations.

\subsection{TDE Flares}

When stars are tidally disrupted by a low-mass black hole ($M_{\rm{BH}} \lesssim 10^8 M_{\odot}$), electromagnetic radiation is produced from the resulting accretion event. TDEs offer rare glimpses into the otherwise inactive black hole population. One of the most promising pieces of evidence for an IMBH comes from disk spectral fitting of the X-ray TDE in a tidally-stripped satellite galaxy discovered by \citep{Lin2018}. Systematic searches for TDEs in UV and optical surveys have been very successful in identifying TDEs with masses of $M_{\rm{BH}} \sim 10^{5-7} M_\odot$ (see for e.g. \citep{Wevers2017,Gezari2021,Hammerstein2023}). \cite{Angus2022} report a fast-rising TDE in a nearby dwarf galaxy. Fast-rising TDEs might be strong IMBH candidates if the light curve rise time correlates with black hole mass \citep{Mockler2019}. TDE black hole masses, however, remain very uncertain, often relying on unreliable light curve modeling or indirect host galaxy scaling relations. A TDE originating from a disrupted white dwarf would be a smoking gun signature of a $M_{\rm{BH}} < 10^5 M_\odot$ IMBH [e.g., \cite{Gezari2021}]. Given the high densities of white dwarfs, they can only be tidally disrupted by IMBHs. Several archival candidates for white dwarf TDEs exist, and rapid spectroscopic characterization of Rubin-discovered TDEs will be critical to confirming their nature \citep{Gomez2023}.

Beyond straightforward TDEs, other transient accretion events might trace the presence of an IMBH. For example, X-ray flares with quasi-periodic eruptions can be modeled as an orbiting IMBH punching through an SMBH accretion disk \citep{Arcodia2021,Chakraborty2025}. Optically selected ambiguous nuclear transients (ANTs) in low-mass galaxies might also be promising in the search for IMBHs. ANTs have features of both changing-state AGNs and TDEs, challenging some of our notions of distinct TDE and AGN classes (e.g., \citep{Hinkle2023}). TDEs from IMBHs are also implicated as a possible explanation for another new type of rarer transient, FBOTs (Fast Blue Optical Transients), which are characterized by a more rapid rise and decay compared to supernovae (SNe); higher luminosities exceeding those of SNe, and bluer spectra that can explain sources like AT2018cow \citep{Kuin+2019}. Additionally, intriguing new classes of transient sources like AT\,2020vdq, that appears to be a repeating partial TDE produced by an IMBH with mass $\sim 10^{5.6}\,{M_{\odot}}$, are also being detected \cite{Somalwar+2025}. High-quality Rubin light curves and triggered multiwavelength follow-up observations will be critical in determining the physical origins of the mechanisms that cause these transient events, which reveal the presence of IMBHs.

\section{Prospects for detecting populations of IMBHs across cosmic time}\label{sec:prosp}

Understanding the demographics of IMBHs requires connecting their observable signatures to both local dwarf galaxies and high-redshift AGN populations. Variability studies offer complementary routes to constrain their abundance and growth histories. IMBHs have distinct observational signatures: in AGNs via their stochastic variability timescales, and TDEs via the light curve shapes. Together, these results frame the prospects for identifying and characterizing IMBH populations across cosmic time.

\begin{figure}[ht]
\centering
\includegraphics[width=0.5\textwidth]{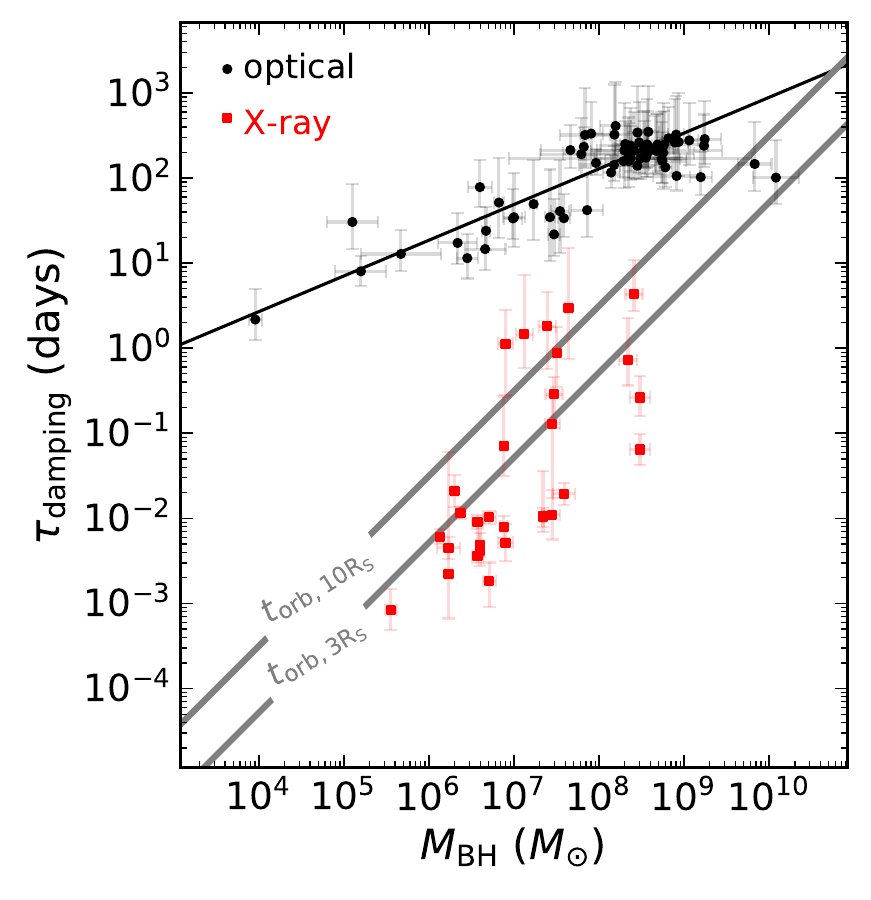}
\caption{\textbf{AGN variability timescales at optical and X-ray wavelengths as functions of SMBH mass.} Red squares show measurements of the break timescale for X-ray variability \cite{Gonzalez-Martin2012} and the black circles are optical measurements. The thick gray lines indicate the orbital timescale at ten and three times the Schwarzschild radius. The error bars are 1$\sigma$ uncertainties. The characteristic variability timescales is $\sim$ hours -- days in the optical or $\sim$ seconds -- minutes in the X-ray for an IMBH. Figure from: \citep{Burke2021science}. }\label{fig:timing}
\end{figure}

\subsection{Wandering black holes}

Cosmological $N$-body simulations show that IMBHs are unlikely to sink to the centers of their host galaxies unless, perhaps, they are affixed to dense nuclear star clusters \citep{Ma2021,Ricarte2021,DiMatteo2023}. \citep{Mezcua2020} selected 37 off-nuclear AGN candidates using integral field unit spectroscopy, 23 of which were missed by integrated emission line diagnostics. \citep{Reines2020} identified 13 probable AGN in $z<0.055$ dwarf ($M_{\star} \leq 3{\times}10^9 M_{\odot}$) galaxies, most of which are spatially offset from the galaxy centers. However, \citep{Eftekhari2020} argue that the off-nuclear sources are likely associated with fast radio bursts. Some IMBHs may be ejected from their galactic centers by gravitational recoil, bringing their accretion disks with them (e.g., \citep{Ward2021}). One possibility is that some IMBHs may masquerade as ultraluminous X-ray sources. Nevertheless, high-resolution difference imaging can pinpoint the position of transient and variable sources, because the change in flux enables their detection against the constant host galaxy background. While contamination from background sources remains a challenge, searches for off-nuclear TDEs and variable AGNs to uncover IMBHs is a relatively unexplored parameter space at the present time. 

\subsection{The Black Hole Masses of $z>0$ Dwarf AGNs}

Beyond the low redshift universe, several authors have pushed the limits of sensitive radio, X-ray, and optical variability surveys to identify dwarf AGNs in deep fields \citep{Mezcua2018,Burke2022des,Zou2023,Mezcua2023}. However, $z>0.1$ optical variability-selected AGNs appear to have overmassive black holes compared to the typical $M_{\rm{BH}}/M_{\star}$ ratios of local AGNs \citep{Burke2024hsc,Liu2025} of $\sim 0.025\%$ \citep{Reines2015}. Part of this is likely due to selection biases (e.g., \citep{Lauer2007,Li2025}), but it is also possible that local AGNs in late-type galaxies have unique growth histories. While identifying lower-mass SMBHs at $z>0$ remains essential for studies of host galaxy-SMBH coevolution and tracking SMBH demographics over cosmic time, the implications of these results suggest that we will have to probe down to even lower stellar masses to identify IMBHs beyond the local Universe. 

\subsection{AGN timing}

The timing properties of AGN variability offer a powerful tool for constraining AGN geometry and estimating black hole masses. Although accretion disks likely have internal processes that produce variability, the standard picture is one where light from a driving X-ray light curve is reprocessed and re-emitted at farther distances in the disk, broad line region, and dusty torus (e.g., \citep{Cackett2007}). The stochastic properties of the light curves correlate with black hole mass (e.g., \citep{Kelly2009,MacLeod2012,Burke2021science,Arevalo2024}), although the origin of the underlying physics is unknown and accretion disks likely produce variability from their own internal processes (e.g., \citep{Hagen2023}). By measuring the lag in multiwavelength light curves, assuming a model-dependent transfer function, one can make further inferences about AGN properties, such as disk sizes and black hole mass (e.g., \citep{Cackett2007,Starkey2017}).

Only about 50 stellar-mass accreting black holes are currently known, providing probes of accretion physics on minute-to-second timescales, but IMBHs will found in much greater numbers. Their variability will be readily captured by LSST, enabling detailed population-level tests of accretion and variability models. For example, the variability timescales in accretion disks around IMBHs are orders of magnitude shorter than in SMBHs, as shown in Figure~\ref{fig:timing} \citep{Burke2021science}. Looking for variability with a characteristic timescale of $\sim$ days could be a straightforward way to identify IMBHs that does not rely on host galaxy stellar mass estimation. This could enable their discovery and eventual use as laboratories for studying accretion on human lifespan compatible timescales. The advent of time-domain fitting using Gaussian Processes has alleviated some of the historical challenges facing optical variability studies, such as irregular light curve sampling, at the cost of assuming a statistical model for the underlying variable process \citep{Kelly2014}. Recent approaches that take advantage of AGN multiband covariance offer the potential to probe timescales shorter than the per-band cadence, reaching hours timescales for LSST light curves [W. Yu et al., in prep].

\subsection{The Black Hole Occupation Fraction}

The fraction of local dwarf galaxies hosting IMBHs is an important constraint on cosmological models of black hole seeding and growth. To obtain meaningful constraints on the occupation fraction derived from incomplete observations of AGNs, we must account for the intrinsic distribution of Eddington ratios of the sample and the detection limit (e.g., \cite{Kelly2007}). Similarly, the occupation fraction inferred from TDEs depends on constraining their Eddington ratios and the completeness-corrected rates; both the rate at which stars are scattered into the loss cone and the fraction of stars that disrupt outside of the event horizon to produce observable flares \citep{Yao2023}. \citep{Miller2015} used a Bayesian linear regression approach to constrain the occupation fraction to $\gtrsim50\%$ at $M_{\star}\sim10^{9} M_{\odot}$ for volume-limited datasets. We recently devised a comprehensive technique to jointly infer the luminosity function and occupation fraction for flux-limited X-ray, radio, and optical variability datasets \citep{Burke2025}. We find a high local occupation fraction of at least 90$\%$ at $M_{\star} \sim 10^8 M_{\odot}$. The occupation fraction can be converted into a black hole mass function by convolving the product of the galaxy stellar mass function and occupation fraction with the $M_{\rm{BH}} - M_{\star}$ relation \citep{Kelly2012,Gallo2019}. The AGN luminosity can be estimated from optical light curve properties, so statistics of Rubin-selected variable AGN light curves plus host galaxy stellar mass estimates can be used to infer the occupation fraction. Currently, there is no clear consensus on the occupation fraction in local dwarf galaxies, for instance, \citep{Zou2025} report finding a much lower occupation fraction using a more flexible model compared to the estimate of \cite{Burke2025}. Therefore, at the present time, the occupation fraction, a clean and useful metric, remains poorly constrained beyond $z=0$ \citep{Ricarte2018,Burke2025}. Similar ``missing data'' statistical approaches could be independently applied to infer the occupation fraction, occurrence rates, and intrinsic luminosity function from TDE light curves. However, these approaches rely on several assumptions, such as the Eddington ratio distribution function, which remains very uncertain at low masses and accretion rates.

\section{Challenges with variability} \label{sec:challenges}

Variability is playing an increasingly important role in expanding the number of IMBHs and opening a new window into accretion physics. As with all methods for identifying IMBHs, variability also has its own drawbacks. Accreting IMBHs have spectral energy distributions that peak in the UV or soft X-rays. IMBH TDEs fade rapidly, where time-domain coverage is limited, while accreting IMBHs show modest amplitudes and day-to-week timescales that surveys undersample and host galaxies dilute \citep{Burke2023demo}. Distinguishing these signals from stochastic AGN variability or supernovae requires high-cadence, multiband data. Once candidates are found, confirming an IMBH demands multi-wavelength mass estimates, making variability-based discovery highly sensitive to survey cadence, selection effects, and limited follow-up resources. Nevertheless, we are optimistic that upcoming time-domain surveys like those on Rubin and the upcoming ULTRASAT space mission \citep{Shvartzvald2024} are poised to advance discovery through improved cadence, depth, and UV sensitivity.

\section{Conclusions} \label{sec:concl}

The past few years have seen huge advances in the detection of numbers of TDEs and AGNs identified in dwarf galaxies with $M_{\rm{BH}} \sim 10^{4-7} M_{\odot}$ to the point where we are beginning to perform demographic studies \citep{Burke2023demo}; document their local environments \citep{Kristensen2020} and map out their accretion properties \citep{Birchall2020}.  Despite this progress, IMBHs still remain largely elusive as a population. Rubin is expected to discover about a thousand accreting IMBHs with $M_{\rm{BH}} <10^6 M_{\odot}$ in dwarf AGNs \citep{Burke2023demo}, a major leap in our census. Similarly, Rubin will expand the present-day detection rate of tens of TDEs discovered per year to hundreds to a few thousand per year \citep{Bricman2020}. In the local Universe, dynamical IMBH signatures will be revealed by the new planned 30-m class telescopes. The LISA mission will open a new window into the IMBH population in-between the stellar mass black holes revealed by LIGO/VIRGO/KAGRA and SMBHs from pulsar timing arrays. Variability studies are, however, independent of the star formation properties of host galaxies, and instead serve as a close tracer of black hole accretion that can be leveraged to derive IMBH masses. Multiwavelength accretion signatures are an independent and complementary approach to future observational discoveries with time-domain surveys with Rubin, the Next Generation Very Large Array, and future X-ray missions. These electromagnetic observations are essential for constraining the accretion properties, local environments, and obtaining a more complete census of the IMBH population. 

\bmhead{Data Availability}

All data-sets presented in this manuscript are public, and the authors would be happy to share their compilation upon request. 

\bmhead{Acknowledgements}



C.J.B. is supported by an NSF Astronomy and Astrophysics Postdoctoral Fellowship under award AST-2303803. This material is based on work supported by the National Science Foundation under Award No. 2303803. This research award to NSF is partially funded by a generous gift of Charles Simonyi to the NSF Division of Astronomical Sciences. The award is made in recognition of significant contributions to Rubin Observatory’s Legacy Survey of Space and Time. 
P.N. acknowledges support from the Gordon and Betty Moore Foundation and the John Templeton Foundation, which fund the Black Hole Initiative (BHI) at Harvard University, where she is a PI.

\section*{Declarations}



\subsection{Conflict of interest/Competing interests}
The authors declare no competing interests.

\subsection{Author Contributions}

CJB: Conceptualization, Data Compilation \& Visualization, Execution and Writing-Original Draft, Review as well as Editing; PN: Motivation, Conceptualization, Execution, and Writing-Original Draft, Review as well as Editing.

\bibliography{sn-bibliography}

\end{document}